\documentstyle [twocolumn,prc,aps,epsfig]{revtex}
\newcommand{\be}{\begin{equation}}
\newcommand{\ee}{\end{equation}}
\newcommand{\bd}{\begin{displaymath}}
\newcommand{\ed}{\end{displaymath}}
\newcommand{\ba}{\begin{eqnarray}}
\newcommand{\ea}{\end{eqnarray}}

\newcommand{\ave}[1]{\langle {#1} \rangle}
\begin {document}
\twocolumn[\hsize\textwidth\columnwidth\hsize\csname @twocolumnfalse\endcsname

\title{Dropping $\sigma$-Meson Mass and 
In-Medium S-wave $\pi$-$\pi$ \\ Correlations}

\author{Z. Aouissat$^1$, G. Chanfray$^2$, P. Schuck$^3$, J. Wambach$^1$}

\address{
$^1$ {\small{\it IKP, Technische Universit\"at 
Darmstadt, Schlo{\ss}gartenstra{\ss}e 9, 64289 Darmstadt, Germany.}}\\
$^2$ {\small{\it  IPN Lyon. 43 Bd. du 11 Novembre 1918, F69622 Villeurbanne
 C\'edex, France. }}\\
$^3$ {\small{\it  ISN, Universite Joseph Fourier, CNRS-IN2P3,
 53 avenue des Martyrs,
F-38026 Grenoble C\'edex, France.}}
} 
\date{ \today }
\maketitle

\begin{abstract}
The influence of a dropping $\sigma$-meson mass on previously 
calculated in-medium $\pi\pi$ correlations in the 
$J=I=0$ ($\sigma$-meson) channel \cite{borm, arcsw} is investigated. It is found that the 
invariant-mass distribution around the vacuum threshold
experiences a further strong enhancement with respect to standard many-body effects.
The relevance of this result for the explanation of recent  
$A(\pi,2\pi)X$ data is pointed out.

\end{abstract}
\vspace{0.6cm}
]





In medium s-wave pion-pion correlations have recently attracted much attention
both on the theoretical \cite{snc,borm,arcsw,rd,cov,vicent,hat}
and experimental \cite{bonu} sides.
These studies are of relevance for the behavior of the in-medium
chiral condensate and its fluctuations with increasing density \cite{hat}.
In earlier studies we have shown that standard p-wave coupling of the pion 
to $\Delta$-h and p-h configurations 
induces a strong enhancement of the $\pi\pi$ invariant-mass distribution around the 
$2m_{\pi}$ threshold \cite{borm,arcsw}, thus signalling increased
fluctuations in the $\sigma$-channel. This fact was independently confirmed 
in \cite{cov}. It has been argued in \cite{borm,rd}
that this effect could possibly explain the $A(\pi,2\pi)$ knockout reaction
data from the CHAOS collaboration \cite{bonu}. 
More recently Vicente Vacas and Oset  \cite{vicent} have claimed that the theory 
underestimates the experimentally found $\pi-\pi$ mass enhancement. This claim may be
partly questioned, since the reaction theory calls for a calculation with a finite total 
three momentum of the in-medium pion pairs \footnote{It was shown in \cite{borm} that
considering the finite three momenta of the pion pair  
may further increase the effect of the in-medium pion-pion final-state interaction on
the final $\pi^+\pi^-$ invariant mass-distribution  at threshold.
Furthermore, at finite total three momenta of the pair, the sigma-meson couples directly 
to particle-hole excitations as well, which  results into  more strength enhancement at threshold,
as shown in \cite{hat}. }.
 On the other hand Hatsuda et al. \cite{hat} argued that the partial restoration of 
chiral symmetry in nuclear matter,  which leads to a dropping of the $\sigma$-meson mass  
\cite{br}, 
induces similar effects as the standard
many-body correlation mentioned above. It is therefore natural to study the
combination of both effects. This is the objective of the present note.

As a model for $\pi\pi$ scattering we consider the linear sigma model treated in 
leading order of the $1/N$-expansion \cite{asw}.
The scattering matrix can then be cast in the following form
\begin{eqnarray}
T_{ab,cd}(s) \,&=& \delta_{ab}\delta_{cd} 
\frac{D_{\pi}^{-1}(s) - D_{\sigma}^{-1}(s)}{3\ave{\sigma}^2}
\, \frac{D_{\sigma}(s)}{D_{\pi}(s)}~,
\label{eq1}
\end{eqnarray}
where $s$ is the Mandelstam variable. In Eq.~(\ref{eq1}) $D_{\pi}(s)$ and $D_{\sigma}(s)$ 
are respectively the full pion and sigma propagators,
while $\ave{\sigma}$ is the sigma condensate. The expression in Eq.~(\ref{eq1}) 
reduces  in fact, in the soft pion limit, to a Ward identity which links the $\pi\pi$ 
four-point function to the $\pi$- and $\sigma$ two-point functions as well
as to the $\sigma$ one-point function.
To this order, the pion propagator and the sigma-condensate are obtained
from the Hartree-Bogoliubov (HB) approximation \cite{asw}. 
In terms of the pion-mass $m_{\pi}$ and the decay constant 
$f_{\pi}$, they are given by  
\begin{equation}
D_{\pi}(s) = \frac{1}{s - m_{\pi}^2}, \quad\quad f_{\pi} \,=\,\sqrt{3} \ave{\sigma}.
\label{eq2}
\end{equation} 
The sigma meson, on the other hand,
is obtained from the Random Phase Approximation (RPA) involving
$\pi$-$\pi$ scattering \cite{asw} and reads    
\begin{equation}
D_{\sigma}(s) \,=\, \left[{ s\,-\, m_{\sigma}^2
\,-\, \frac{2 \lambda^4 \ave{\sigma}^2\,{\Sigma}_{\pi\pi}(s)}
{ 1\,-\,  \lambda^2 {\Sigma}_{\pi\pi}(s)}}\right]^{-1}~,
 \label{eq3}
\end{equation}
where ${\Sigma}_{\pi\pi}(s)$ is the $\pi\pi$ self-energy regularized by means 
of a form factor which is used as a fit function \cite{borm} and allows 
to reproduce the experimental $\pi\pi$ phase shifts.   
The coupling constant $\lambda^2$ denotes the bare quartic coupling of the 
linear $\sigma$-model, 
related to the mean-field pion mass $m_{\pi}$,  sigma mass  $m_{\sigma}$, 
and the condensate $\ave{\sigma}$ via the mean-field saturated  Ward identity
 
\begin{equation}
m_{\sigma}^2 = m_{\pi}^2 + 2 \lambda^2\ave{\sigma}^2. 
 \label{eq4}
\end{equation}
It is clear from what was said above that the $\sigma$-meson
propagator in this approach is correctly defined, since it satisfies a whole 
hierarchy of Ward identities.

In cold nuclear matter the pion is dominantly coupled to
$\Delta$-h,  p-h, as well as to the 2p-2h excitations which, on the other hand, 
are renormalized by means of repulsive nuclear short-range correlations,   
(see \cite{arcsw} for details). 
Since the pion is a (near) Goldstone mode,  its in-medium  s-wave renormalization 
does not induce considerable changes. The sigma meson, on the other hand,
is not protected against an important s-wave 
renormalization from chiral symmetry.  
Therefore, following a very economical procedure, we 
extract an approximate density dependence of the mean-field sigma meson 
mass by taking into account the density dependence of
the condensate. From eq.(\ref{eq4}) it is clear that the density dependence
of the sigma-meson is essentially dictated by the density dependence of the 
condensate.

\begin{figure}
\centerline{ 
\epsfig{file=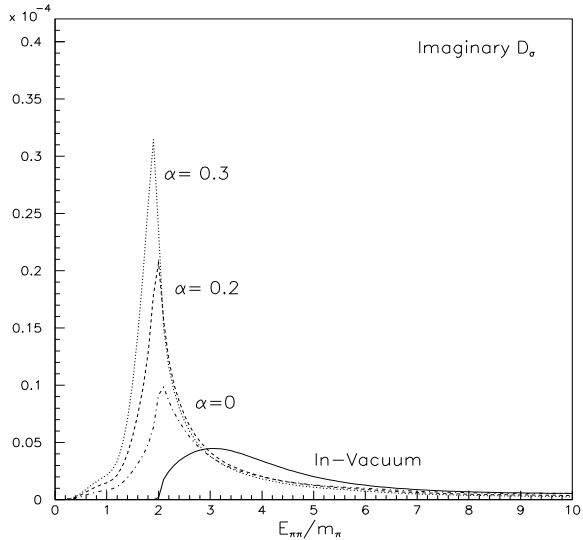,width=8.6cm,height=8.cm,angle=0}}
\caption[fig1]{\small
Results for the imaginary part of the in-medium sigma-meson propagator. 
Except for the vacuum case (full line curve) 
the remaining in-medium curves are 
computed at normal nuclear matter density. 
The dashed-dotted curve is for $\alpha=0$, dashed 
for $\alpha=0.2$ and the dotted for $\alpha=0.3$.
\label{fig1.} }
\end{figure}

For densities below and around nuclear saturation density $\rho_0$ we take 
for the in-medium $sigma$-meson  mass the simple ansatz (see also \cite{hat})
\begin{equation} 
m_{\sigma}(\rho)= m_{\sigma}(1 -\alpha \frac{\rho}{\rho_0})
\label{eq5}
\end{equation} 
where $\rho$ is the nuclear matter 
density and $m_{\sigma}$ is the vacuum  $\sigma$-meson mass.    
The parameter $\alpha$ can be estimated from model calculations or QCD sum rules
and lies in the range from 0.2 to 0.3. These are the values
which we also will use in this work. 
 
The result for the sigma-meson mass distribution $Im D_{\sigma}(E_{\pi\pi})$, as
calculated from Eq.~(3) by using the in-medium mass (\ref{eq5}), is shown in 
Fig.~2 for various densities. One sees that, as density increases, a strong downward 
shift of the sigma-mass distribution occurs.
The enhancement at low energies is strongly reinforced as
the in-medium $\sigma$-meson mass is included. For $\alpha=0.2$ and 
$\alpha=0.3$ the peak height is increased by a factor 2 and 3 respectively.
Similarly for the T-matrix, a sizeable effect can be noticed in its imaginary
part. There is therefore a large flexibility
 to explain the findings of
CAHOS collaboration\footnote{ Comparing the curve for $\alpha=0.2$, for instance,  and the
curves of figure 4 (from ref.~\cite{borm}),   which where used to compute the
$\pi^+ \pi^-$  mass distribution (figure 5 of \cite{borm}),  one realizes  that there is indeed 
enough strength at threshold to reproduce the experiemtal data.}.   
 Work in this direction is in progress.

\begin{figure}
\centerline{ 
\epsfig{file=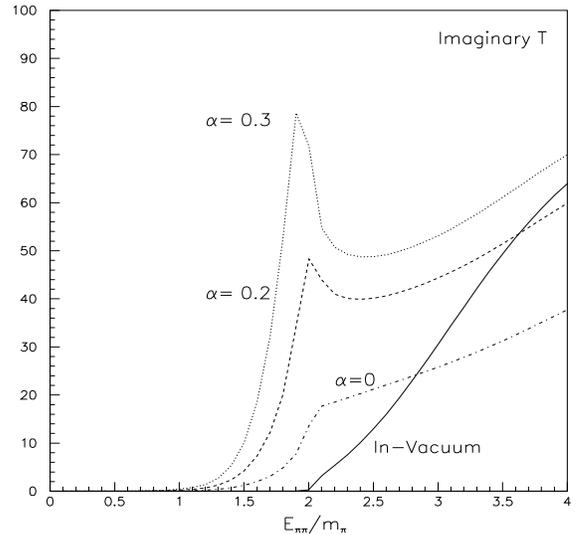,width=8.6cm,height=8.cm,angle=0}}
\caption[fig1]{\small
Results for the imaginary part of the in-medium 
T-matrix for $\pi\pi$ scattering. 
Except for the vacuum case (full line curve) 
the remaining in-medium curves are 
computed at normal nuclear matter density. 
 The dashed-dotted curve is for $\alpha=0$, the dashed 
for $\alpha=0.2$ and the dotted for $\alpha=0.3$.
\label{fig2.} }
\end{figure}

These findings call for some comments. It is clear that vertex 
corrections, usually a source of repulsion and not taken into account in this work, could 
weaken the effects. In particular the incidence 
of the nuclear density on the coupling constants should be considered. This  seems, however, 
to be of minor importance as was recently shown by Chanfray et al.~\cite{dd}. More care should 
also be taken in properly incorporating Pauli-blocking  when renormalizing
the pion pairs in matter, although preliminary investigations \cite{thesis}
have shown that this effect is weak.

In conclusion we have shown that a dropping sigma-meson mass, linked to 
the partial restoration of chiral symmetry in nuclear matter, 
further enhances the build up of previously found $\pi\pi$ strength 
in the $I=J=0$ channel.
Further studies are necessary to show how precisely this is linked to
the recent findings by Bonutti et al.~\cite{bonu}.

\end{document}